\documentclass{article}

\usepackage{arxiv}

\usepackage[utf8]{inputenc} 
\usepackage[T1]{fontenc}    
\usepackage{hyperref}       
\usepackage{url}            
\usepackage{booktabs}       
\usepackage{amsfonts}       
\usepackage{nicefrac}       
\usepackage{microtype}      
\usepackage{lipsum}		
\usepackage{graphicx}
\usepackage{natbib}
\usepackage{doi}

\usepackage{multirow}
\usepackage{stfloats}
\usepackage{amsmath,amssymb,amsfonts}
\usepackage{algorithmic}
\usepackage{graphicx}
\usepackage{textcomp}

\title{HiFuse: Hierarchical Multi-Scale Feature Fusion Network for Medical Image Classification}

\author{Xiangzuo Huo\thanks{Equal contribution} \\
	Xinjiang University\\
	\texttt{huoxiangzuo@163.com} \\
	\And
	Gang Sun$^*$\\
	The Affiliated Cancer Hospital of \\
	Xinjiang Medical University\\
	\texttt{sung853219@163.com}
    \And
    Shengwei Tian\thanks{Corresponding author}\\
    Xinjiang University\\
    \texttt{tianshengwei@163.com}
    \And Yan Wang$^\dag$ \\
	The Affiliated Cancer Hospital of \\
	Xinjiang Medical University\\
    \And Long Yu \\
    Xinjiang University\\
    \And Jun Long \\
    Central South University \\
    \And Wendong Zhang \\
    Xinjiang University\\
    \And  Aolun Li \\
    Xinjiang University\\
}

\renewcommand{\shorttitle}{\textit{arXiv} Template}

\begin{document}
\maketitle

\begin{abstract}
Medical image classification has developed rapidly under the impetus of the convolutional neural network (CNN). Due to the fixed size of the receptive field of the convolution kernel, it is difficult to capture the global features of medical images. Although the self-attention-based Transformer can model long-range dependencies, it has high computational complexity and lacks local inductive bias. Much research has demonstrated that global and local features are crucial for image classification. However, medical images have a lot of noisy, scattered features, intra-class variation, and inter-class similarities. This paper proposes a three-branch hierarchical multi-scale feature fusion network structure termed as HiFuse for medical image classification as a new method. It can fuse the advantages of Transformer and CNN from multi-scale hierarchies without destroying the respective modeling so as to improve the classification accuracy of various medical images. A parallel hierarchy of local and global feature blocks is designed to efficiently extract local features and global representations at various semantic scales, with the flexibility to model at different scales and linear computational complexity relevant to image size. Moreover, an adaptive hierarchical feature fusion block (HFF block) is designed to utilize the features obtained at different hierarchical levels comprehensively. The HFF block contains spatial attention, channel attention, residual inverted MLP, and shortcut to adaptively fuse semantic information between various scale features of each branch. The accuracy of our proposed model on the ISIC2018 dataset is 7.6\% higher than baseline, 21.5\% on the Covid-19 dataset, and 10.4\% on the Kvasir dataset. Compared with other advanced models, the HiFuse model performs the best. Our code is open-source and available from \underline {https://github.com/huoxiangzuo/HiFuse}.
\end{abstract}

\keywords{Medical Image Classification\and Feature Fusion\and Swin Transformer\and Hybrid Network\and Muti-scale Feature}

\section{Introduction}
\label{sec:introduction}

Medical image classification is an important task in computer-aided diagnosis, medical image retrieval, and mining. In recent years, convolutional neural networks have achieved outstanding performances in many medical image classification tasks\cite{Koitka2016,Xu2014,Shen2017,Esteva2017,Personnaz1986,Kumar2016,Yu2017}. However, medical images have high similarity and detail diversity in imaging modalities and clinicopathology, resulting in significant intra-class variation and inter-class similarity, requiring modeling of global semantic information, so medical image classification remains challenging.

Transformer\cite{Vaswani2017} was originally used for modeling sequence-to-sequence prediction in natural language processing (NLP) tasks. Now Transformer has also attracted much attention in the computer vision community, ViT\cite{Dosovitskiy2020} by segmenting each image into patches with positional embeddings. Sequences of tokens are constructed, and a cascaded transformer block is applied to extract parameterized vectors as visual representations that model global semantic information through complex spatial transformations and long-range feature dependencies. Due to the lack of local spatial feature details. Li et al.\cite{Yuan2021} proposed to utilize CNN feature maps as input tokens to capture feature neighborhood information. However, they model the image as a one-dimensional sequence of tokens, ignoring the local inductive bias of the image, which affects the convergence speed and performance of the model.

Recent studies, such as ViTAE\cite{Xu2021}, StoHisNet\cite{Fu2022}, Transfuse\cite{Zhang2021}, CMT\cite{Guo2022a}, Comformer\cite{Peng2021} and so on, which solve the above problems to a certain extent by Fusion of the features extracted by the convolution and self-attention mechanism\cite{Vaswani2017}. Different from the above methods, to further exploit the advantages of CNN and Transformer in the medical field, a  three-branch parallel hierarchical fusion network structure termed as HiFuse is proposed as a new method for medical image classification to improve the classification accuracy of various medical images. Taking ConvNext\cite{Liu2022} as our baseline and inspired by ResNet\cite{he2016deep} and Swin Transformer\cite{Liu2021}, we designed new local and global feature blocks. HFF blocks fuse local and global representations of various semantic scales. This fusion process can fully mine the deep-shallow and global-local features of the lesion area in the medical image classification task.

HiFuse has the following advantages:

1. Combining the advantages of CNN and Transformer, a parallel framework of Local and Global feature blocks is designed to efficiently capture local spatial context features and global semantic information representation of features at different scales, respectively. In addition, HiFuse does not need to build a very deep network to achieve good results, effectively avoiding the problems of gradient disappearance and loss of feature information.

2. An adaptive hierarchical feature fusion block (HFF block) is designed, which contains spatial attention, channel attention, residual
inverted MLP, and shortcut connection to adaptively fuse semantic information between different scale features of each branch. 

3. The proposed HiFuse model achieves relatively good results on ISIC2018, Covid-19, and Kvasir datasets.

\begin{figure}[t]
\centering
\includegraphics[width=6 in]{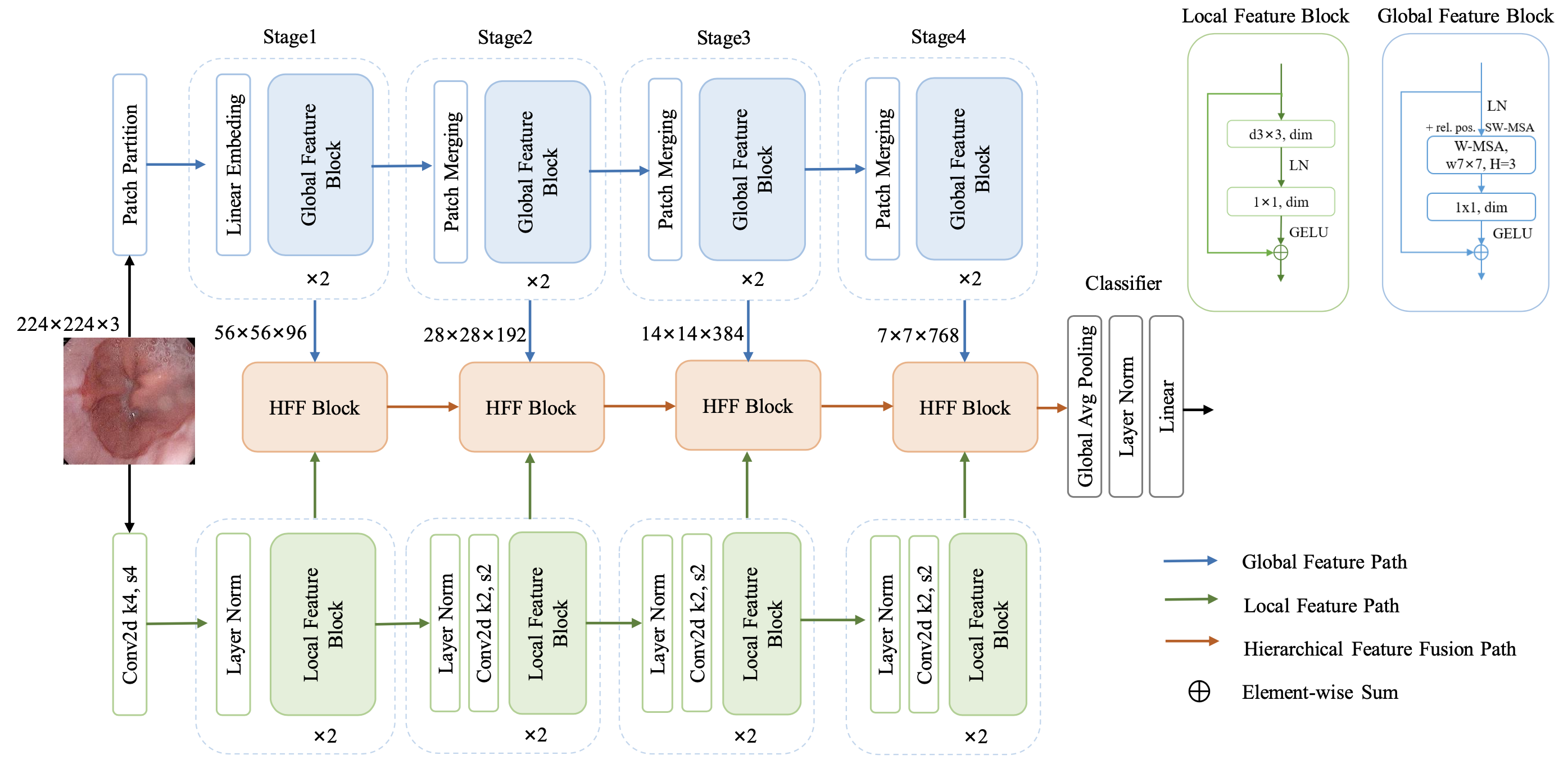}
\caption{Overall structure of the HiFuse model.}
\label{fig1}
\end{figure}

\section{Related Works}
Traditional medical image classification methods employ color, texture, shape, and combined descriptors. Baloch et al.\cite{Baloch2007} proposed a flexible skew-symmetric shape model to learn to capture latent changes within a certain neighborhood. Song et al.\cite{Song2013} proposed a novel texture descriptor that represents texture features by integrating multi-scale Gabor transform and local binarized histograms to classify lung tissue. Koitka et al. \cite{Koitka2016} manually extracted visual descriptors and used them for medical image classification.

Medical image classification work based on deep learning has emerged in recent years. The deep convolutional neural network (DCNN) method that has greatly improved the classification accuracy and reduced the waste of resources for manual feature extraction is gradually applied clinical auxiliary diagnosis. Xu et al. \cite{Xu2014} used DCNN to extract features from histopathological images of colon cancer and achieved good classification results. Shen et al. \cite{Shen2017} proposed a multi-scale crop pooling strategy for DCNN to capture lung nodule classification features on chest CT images.

Esteva et al. \cite{Esteva2017} used a model trained on 129,450 clinical images to diagnose the most common and deadly skin cancers and achieved performance matching 21 dermatologists. Koitka et al. \cite{Koitka2016} custom-train the output of the last fully connected layer in a pre-trained ResNet-152 model, illustrating that transfer learning can achieve good results in medical image classification. Kumar et al. \cite{Kumar2016} integrated two different pretrained DCNN architectures and combined them into a more powerful classifier. Cheng et al. \cite{Cheng2022} proposed a modular group attention block that captures feature correlations in medical images' channel and space dimensions. However, these models cannot collect sufficient contextual information, and global semantic information features are equally important in high-resolution medical images.

Transformer\cite{Vaswani2017} was originally used in NLP. It extracts intrinsic properties through a self-attention method. ViT \cite{Dosovitskiy2020}, as a pioneer work, verifies the feasibility of pure Transformer architecture in computer vision tasks, and is gradually used for image classification \cite{Yuan2021,Wu2020,Touvron2021}, object detection\cite{Carion2020,Zhu2020,Beal2020}, semantic segmentation\cite{Zheng2021,Zhang2021,Gao2021}, image enhancement\cite{Chen2021} and image generation\cite{Wan2021}. Researchers \cite{Valanarasu2021,He2021,Jiang2021,Wang2021} have tried various approaches to make transformers more successful in computer vision. However, the self-attention mechanism in the visual Transformer often ignores local feature details, and for weak local features, it is not easy to distinguish the object from the background.

To address the lack of local features, DeiT\cite{Touvron2021} proposed using distilled tokens to transfer the CNN-based features to the visual Transformer. T2TViT\cite{Yuan2021} proposed using a tokenization module recursively reorganizing images to consider adjacent pixels' tokens. The models, such as VitAE, StoHis, TransFuse, CMT, Conformer and so on,\cite{Xu2021,Fu2022,Zhang2021,Guo2022a,Peng2021} not only  inherit the structural advantages of CNN and Transformer, but also verify that the coupling of local features and global representation can significantly enhance Transformer discriminability of weak local features. The above models perform well on natural datasets such as ImageNet and various downstream tasks, however, when applied to the medical image domain, the results are unsatisfactory. Because the datasets of medical images are insufficient, pathological features are more scattered and difficult to discover than those of ordinary images. So, we decide to take full advantage of deep-shallow and global-local features and a wider fusion network to fuse them. Our proposed HiFuse model defines a three-branch hierarchical parallel fusion structure, combines local with global feature blocks, and designs a hierarchical feature fusion block (HFF block) to fuse these features and keep local and global branches undisturbed. HiFuse inherits not only the advantages of CNN and Transformer but also local features and global representations that are coupled at different scales.

\begin{table}[t]
\centering
\caption{HiFuse specific parameters.}
\begin{tabular}{cc|cccrcc}
\hline
\multicolumn{1}{c|}{\textbf{stage}}        & \textbf{output size}       & \multicolumn{2}{c|}{\textbf{Local Branch}}                                                                                                    & \multicolumn{2}{c|}{\textbf{HFF Branch}}                                                       & \multicolumn{2}{c}{\textbf{Global Branch}}                                                                                                             \\ \hline
\multicolumn{1}{c|}{\multirow{2}{*}{stem}} & \multirow{2}{*}{56×56, 96} & \multicolumn{2}{c|}{\multirow{2}{*}{4×4, 96, stride 4}}                                                                                       & \multicolumn{2}{c|}{\multirow{2}{*}{-}}                                                       & \multicolumn{2}{c}{\multirow{2}{*}{4×4, 96, stride 4}}                                                                                                 \\
\multicolumn{1}{c|}{}                      &                            & \multicolumn{2}{c|}{}                                                                                                                         & \multicolumn{2}{c|}{}                                                                         & \multicolumn{2}{c}{}                                                                                                                                   \\ \hline
\multicolumn{1}{c|}{\multirow{2}{*}{1}}    & \multirow{2}{*}{56×56}     & \multicolumn{1}{c|}{\multirow{2}{*}{\begin{tabular}[c]{@{}c@{}}d3×3, 96\\ 1×1, 96\end{tabular}}}   & \multicolumn{1}{c|}{\multirow{2}{*}{×2}} & \multicolumn{1}{l|}{→spatial attention}        & \multicolumn{1}{r|}{channel attention←}      & \multicolumn{1}{c|}{\multirow{2}{*}{\begin{tabular}[c]{@{}c@{}}MSA, w7×7, head 3, \\ rel. pos. SW-MSA\\ 1×1, 96\end{tabular}}}   & \multirow{2}{*}{×2} \\ \cline{5-6}
\multicolumn{1}{c|}{}                      &                            & \multicolumn{1}{c|}{}                                                                              & \multicolumn{1}{c|}{}                    & \multicolumn{2}{c|}{\begin{tabular}[c]{@{}c@{}}d3×3, 96\\ 1×1, 384\\ 1×1, 96\end{tabular}}    & \multicolumn{1}{c|}{}                                                                                                            &                     \\ \hline
\multicolumn{1}{c|}{\multirow{2}{*}{2}}    & \multirow{2}{*}{28×28}     & \multicolumn{1}{c|}{\multirow{2}{*}{\begin{tabular}[c]{@{}c@{}}d3×3, 192\\ 1×1, 192\end{tabular}}} & \multicolumn{1}{c|}{\multirow{2}{*}{×2}} & \multicolumn{1}{l|}{→spatial attention}        & \multicolumn{1}{r|}{channel attention←}      & \multicolumn{1}{c|}{\multirow{2}{*}{\begin{tabular}[c]{@{}c@{}}MSA, w7×7, head 6, \\ rel. pos. SW-MSA\\ 1×1, 192\end{tabular}}}  & \multirow{2}{*}{×2} \\ \cline{5-6}
\multicolumn{1}{c|}{}                      &                            & \multicolumn{1}{c|}{}                                                                              & \multicolumn{1}{c|}{}                    & \multicolumn{2}{c|}{\begin{tabular}[c]{@{}c@{}}d3×3, 192\\ 1×1, 768\\ 1×1, 192\end{tabular}}  & \multicolumn{1}{c|}{}                                                                                                            &                     \\ \hline
\multicolumn{1}{c|}{\multirow{2}{*}{3}}    & \multirow{2}{*}{14×14}     & \multicolumn{1}{c|}{\multirow{2}{*}{\begin{tabular}[c]{@{}c@{}}d3×3, 384\\ 1×1, 384\end{tabular}}} & \multicolumn{1}{c|}{\multirow{2}{*}{×2}} & \multicolumn{1}{l|}{→spatial attention}        & \multicolumn{1}{r|}{channel attention←}      & \multicolumn{1}{c|}{\multirow{2}{*}{\begin{tabular}[c]{@{}c@{}}MSA, w7×7, head 12, \\ rel. pos. SW-MSA\\ 1×1, 384\end{tabular}}} & \multirow{2}{*}{×2} \\ \cline{5-6}
\multicolumn{1}{c|}{}                      &                            & \multicolumn{1}{c|}{}                                                                              & \multicolumn{1}{c|}{}                    & \multicolumn{2}{c|}{\begin{tabular}[c]{@{}c@{}}d3×3, 384\\ 1×1, 1536\\ 1×1, 384\end{tabular}} & \multicolumn{1}{c|}{}                                                                                                            &                     \\ \hline
\multicolumn{1}{c|}{\multirow{2}{*}{4}}    & \multirow{2}{*}{7×7}       & \multicolumn{1}{c|}{\multirow{2}{*}{\begin{tabular}[c]{@{}c@{}}d3×3, 768\\ 1×1, 768\end{tabular}}} & \multicolumn{1}{c|}{\multirow{2}{*}{×2}} & \multicolumn{1}{l|}{→spatial attention}        & \multicolumn{1}{r|}{channel attention←}      & \multicolumn{1}{c|}{\multirow{2}{*}{\begin{tabular}[c]{@{}c@{}}MSA, w7×7, head 24, \\ rel. pos. SW-MSA\\ 1×1, 768\end{tabular}}} & \multirow{2}{*}{×2} \\ \cline{5-6}
\multicolumn{1}{c|}{}                      &                            & \multicolumn{1}{c|}{}                                                                              & \multicolumn{1}{c|}{}                    & \multicolumn{2}{c|}{\begin{tabular}[c]{@{}c@{}}d3×3, 768\\ 1×1, 3072\\ 1×1, 768\end{tabular}} & \multicolumn{1}{c|}{}                                                                                                            &                     \\ \hline
\multicolumn{2}{c|}{Parameters}                                         & \multicolumn{6}{c}{\textbf{82.49 M}}                                                                                                                                                                                                                                                                                                                                                                   \\ \hline
\multicolumn{2}{c|}{FLOPs}                                              & \multicolumn{6}{c}{\textbf{8.13 G}}                                                                                                                                                                                                                                                                                                                                                                    \\ \hline
\end{tabular}
\label{table1}
\end{table}

\begin{figure}[t]
\centering
\includegraphics[width=6 in]{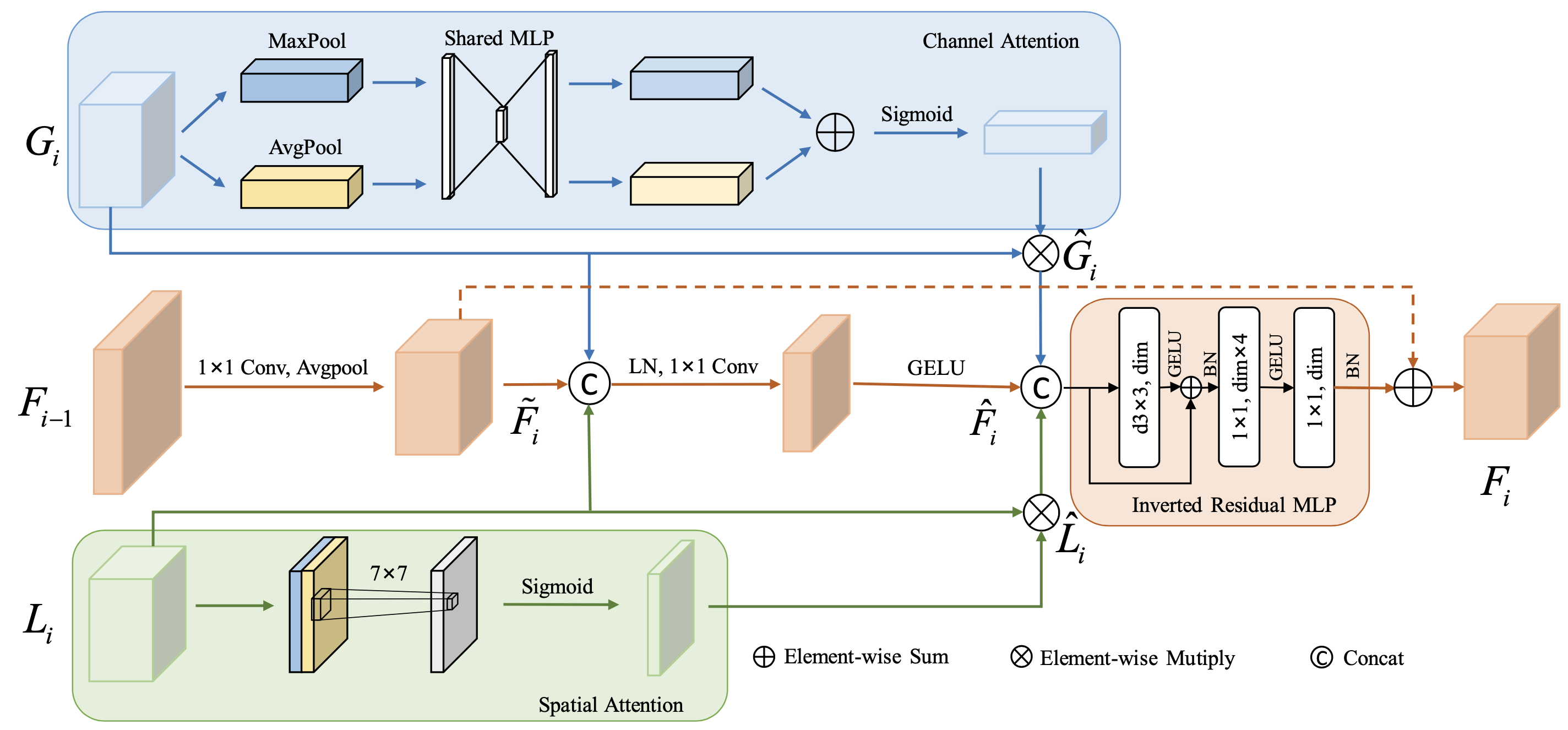}
\caption{HFF block detail display.}
\label{fig2}
\end{figure}

\section{Proposed Method}

\subsection{Overview}
The HiFuse model, as a new medical image classification method, is proposed to effectively obtain local spatial information and global semantic representations of medical images at different scales. We use a parallel structure to extract the global and local information of medical images from the global and local feature blocks, fusing the features of different hierarchies through the HFF block, downsampling step, and finally, obtaining the classification result. In the following sections, we first introduce the overall structure of the HiFuse model, then introduce the global feature block and the local feature block, respectively. We describe the details of the HFF block in Section \ref{HFF}.

\subsection{The Multi-stage Design for HiFuse}
In order to improve the accuracy of the classification model of medical images, it is necessary to fuse local features and global representations from different hierarchical levels. We designed a parallel network structure for hierarchical feature fusion. The overall structure of HiFuse is shown in Figure  \ref{fig1}. The local branch is used to extract the local features of the image, and the global branch is used to extract the global semantic representation of the image. Both branches consist of 4 Stages for feature extraction at different scales. The stem block of the local branch is a 4$\times$4 convolution with stride 4, followed by Layer Norm \cite{Carion2020}. The stem module of the global branch divides the image through the patch partition module. Each 4$\times$4 adjacent pixel is a patch and then flattened in the channel direction. Patch merging changes the output to 2 times the input channel through the linear embedding layer and applies the global feature block for feature transformation. The specific parameters are shown in Table \ref{table1}.

The three-branch parallel structure means that local features and global representations can be preserved to the greatest extent without interfering with each other. Feature maps of different hierarchical levels are constructed through four stages. HFF blocks are used to fuse each stage's local features and global representations and connect the output from the previous stage. The output of local features of each hierarchy of local through spatial attention is combined with the global features of each hierarchy of global through channel attention\cite{Woo2018}. Finally, the combined features are fed to the linear classifiers of Global Average Pooling and Layer Norm for classification. We build different HiFuse variants, HiFuse-Tiny/Small/Base; these variants have different numbers of global and local feature blocks in each stage and build models with different depths to deal with datasets of various sizes. The hyper-parameters of these model variants are:
\begin{itemize}
    \item HiFuse-Tiny: Block numbers = (2, 2, 2, 2)
    \item HiFuse-Small: Block numbers = (2, 2, 6, 2)
    \item HiFuse-Base: Block numbers = (2, 2, 18, 2)
\end{itemize}

\subsection{Global Feature Block}
The imaging methods and clinical pathology of medical images are diverse, with significant intra-class changes and inter-class similarities. The acquisition of global semantic information is very important. Therefore, we introduced the Windows Multi-head Self-Attention (W-MSA) in the global feature extraction branch. W-MSA is the Swin Transformer\cite{Liu2021} first proposed, Compared with the Multi-head Self-Attention (MSA) module in the Transformer, the W-MSA module, which can effectively reduce the amount of computation and divide the feature map into \(M\times M\) sizes. Window one by one, and then perform self-attention on each Window individually. The computational complexity formula is shown in \eqref{eq1}.  
\begin{equation}
\begin{gathered}
\Omega(\text{MSA})=4 h w C^2+2(h w)^2 C\\
\Omega(\text{W-MSA})=4 h w C^2+2 M^2 h w C
\end{gathered}
\label{eq1}
\end{equation}
where \(h\) represents the height of the feature map, \(w\) represents the width of the feature map, \(C\) represents the depth of the feature map, and \(M\) represents the size of each window.

For each stage, by incorporating the patch into the global feature block, the feature map goes through LayerNorm \cite{Ba2016} layer into W-MSA and then through the linear layer with the GELU activation function, as shown in Figure 1. A residual connection is applied after each module, a relative position bias (rel. pos.) is used, and Shift W-MSA is introduced in the next module. This process is depicted in \eqref{eq2}.
\begin{equation}
\begin{gathered}
g_i=f^{1 \times 1}\left(\text{W-MSA}\left(L N\left(G_{i-1}\right)\right)\right)+G_{i-1} \\
G_i=f^{1 \times 1}\left(\text{SW-MSA}\left(L N\left(g_i\right)\right)\right)+g_i
\end{gathered}
\label{eq2}
\end{equation}
where \(G_i\) and \(g_i\) denote the output features of the Shift W-MSA and the W-MSA for the global feature block. \(f^{1\times 1}\) is the convolution operation with a convolution kernel size of \(1\times 1\), which is equivalent to the linear operation. \(LN\) is the LayerNorm operation. Finally, the extracted global features are input into the HFF block.

\subsection{Local Feature Block}
Local spatial features in medical images are also very important. The local feature block, shown in Figure \ref{fig1}, uses a \(3\times 3\) depthwise convolution \cite{Howard2017,Chollet2017}, a special case of grouped convolutions \cite{Xie2017}; the number of groupings is equal to the number of channels. The use of depthwise convolutions effectively reduces the FLOPs of the network. And then, cross-channel information interaction through linear layers obtains good performance in different application scenarios by borrowing the LN and GELU activation functions in the Transformer. Finally, the extracted local features are input into the HFF block. This process is depicted in \eqref{eq3}.
\begin{equation}
L_i=f^{1 \times 1}\left(L N\left(f^{d3 \times 3}(L_{i-1})\right)\right)+L_{i-1}
\label{eq3}
\end{equation}
where \(L_i\) denote the output features of the local feature block. \(f^{d3\times 3}\) is the depthwise convolution operation with a convolution kernel size of \(3\times 3\).

Macroscopically, local and global branch structures are similar, and the design of the same number of channels and hierarchical structure lays the foundation for fusing local and global encoding features of different scales. How to effectively fuse features of different scales in each branch becomes a new problem. To this end, we propose the HFF block.

\begin{figure*}[t]
\centering
\includegraphics[width=6.8 in]{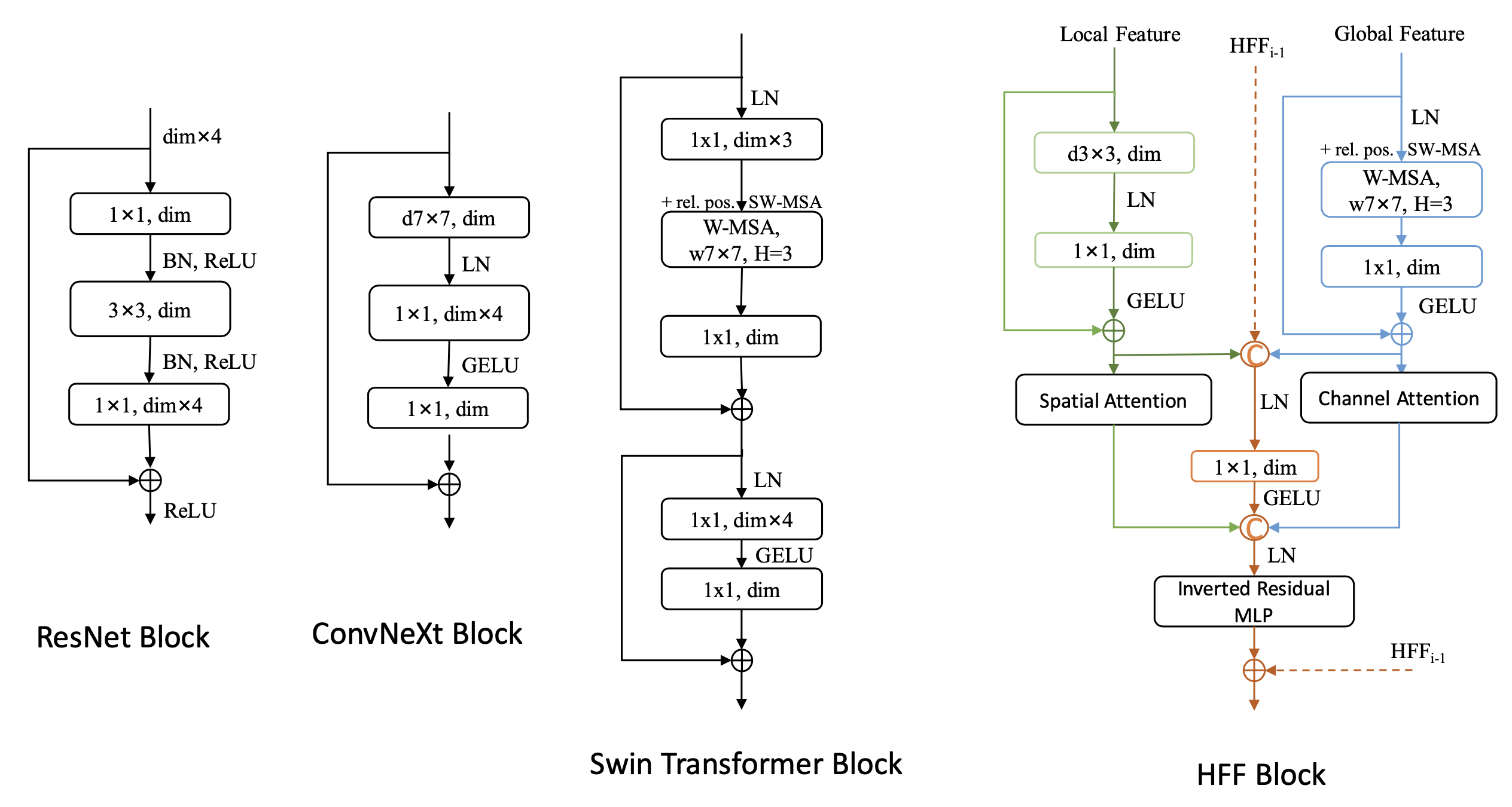}
\caption{Block designs for a ResNet, a ConvNeXt, a Swin Transformer and a HFF.}
\label{fig3}
\end{figure*}

\subsection{HFF Block}
\label{HFF}
Adaptive hierarchical feature fusion block can adaptively fuse local features from different layers, global representations, and semantic information after fusion from the previous hierarchy according to the input features, as shown in Figure \ref{fig2}. Among them, \(G_i\) represents the feature matrix output by the global feature block, \(L_i\) represents the feature matrix output by the local feature block, \(F_{i-1}\) represents the feature matrix output by the previous stage of HFF, and \(F_i\) represents the feature matrix generated by HFF fusion at this stage.

Since the self-attention in the global feature block can capture the spatial and temporal global information\cite{Guo2022} to a certain extent, the HFF block feeds the incoming global features into the channel attention (CA) mechanism, which utilizes the interdependence between the channel maps to improve the feature representation of specific semantics. The local features are input into the spatial attention (SA) mechanism to enhance local details and suppress irrelevant regions. Finally, the results generated by each attention and the fusion path will be feature fusion, and a residual inverted MLP (IRMLP) will be connected. To a certain extent, the problems of gradient vanishing, explosion, and network degradation are prevented, thereby effectively capturing global and local feature information at each hierarchy. For the structure comparison of the structure of ResNet, Swin Transformer, ConvneXt, and our HFF blocks, as shown in the figure \ref{fig3}. This process is depicted in \eqref{eq4}.
\begin{equation}
\begin{gathered}
\text{CA}(x)=\sigma(\operatorname{MLP}(\operatorname{AvgPool}(x))+\operatorname{MLP}(\operatorname{MaxPool}(x))) \\
\text{SA}(x)=\sigma\left(f^{7 \times 7}(\operatorname{Concat}[\operatorname{AvgPool}(x), \operatorname{MaxPool}(x)])\right) \\
\text{IRMLP}(x)=f^{1 \times 1}(f^{1 \times 1}(f^{3 \times 3}\left(LN(x))+LN(x) ) \right) 
\end{gathered}
\label{eq4}
\end{equation}
where \(\sigma\) is the Sigmoid function, \(f^{7 \times 7}\)is the convolution operation with a convolution kernel size of 7$\times $7. The feature fusion operation uses the following formula:
\begin{equation}
\begin{gathered}
\hat{G_i}=\text {CA}\left(G_i\right) \otimes G_i \\
\hat{L_i}=\text {SA}\left(L_1\right) \otimes L_i \\
\tilde{F_i}=\operatorname{Avgpool}\left(f^{1 \times 1}\left(x_{f, i-1}\right)\right) \\
\hat{F_i}=f^{3 \times 3}\left(\operatorname{Concat}\left[G_i, L_i, \tilde{F_{i}}\right]\right)\\
F_i=\text{IRMLP}\left(\operatorname{Concat}\left[\hat{G_i}, \hat{L_i}, \hat{F_i}\right]\right)+\tilde{F_i}
\end{gathered}
\end{equation}
where \(\otimes\) represents element-wise multiply, \(\hat{G_i}\) is generated by the combination of channel attention, \(\hat{L_i}\) is generated by the combination of spatial attention, and
\(\tilde{F_i}\) is generated downsampled by the previous stage of the HFF block.
\(\hat{F_i}\) is the result of global-local features and the fusion of the previous stage. Finally, \(\hat{F_i}\) , \(\hat{G_i}\) and \(\hat{L_i}\) are concatenated and generate feature \(F_i\) through a IRMLP.

\section{Experiments}
\subsection{Dataset}
ISIC2018\cite{Codella2019}: We use the ISIC2018 skin lesion diagnosis dataset task 3. There are 10,015 images in this dataset with seven different categories. They are melanocytic nevi (6705), dermatofibroma (115), melanoma (1113), actinic keratosis (327), benign keratosis (1099), basal cell carcinoma (514), and vascular lesions (142). The size of the images in the dataset is 650 × 450 pixels. We downscale all images to 224 × 224 pixels and according to the division method of chen[], of which 70\% of the samples (7010) are used for training and verification, and the remaining 30\% of the samples (3005) are used for testing.

COVID19-CT\cite{He2020}: This dataset contains 349 COVID-19 positive CT scan images and 397 normal or negative CT scans containing other types of disease. The image sizes in this dataset range from 143×76 to 1637×1225. We scale all images to 224 × 224 pixels, follow the data division method\cite{He2020} and divide the dataset into a ratio of 0.6:0.15:0.25 for training, validation, and testing.

Kvasir\cite{pogorelov2017kvasir}: This dataset includes 4000 endoscopic gastrointestinal
diseases and comprises eight classes, each containing 500 images. The dataset consists of several 
images in each category, including anatomical landmarks
(such as Z-line, pylorus, or cecum) and pathological findings (such as esophagitis, polyps, or ulcerative colitis). Images with
different resolutions from 720 × 576 to 1920 × 1072 pixels are included in the dataset. We downscale all images to 224 × 224 pixels, follow the data division method in literature\cite{pogorelov2017kvasir}, and split the dataset into a 50:50 ratio with 2-fold cross-validation.

\subsection{Performance Metrics}
We choose ACC, F1, Precision, and Recall as classification indicators. These metrics are all calculated based on the confusion matrix. The definitions of the symbols in the confusion matrix are as follows: True Positive (TP), True Negative (TN), False Positive (FP), and False Negative (FN). Therefore, Accuracy (ACC) is calculated by Equation \eqref{eq6} to get the percentage of correctly identified samples.

\begin{equation}
Accuracy=\frac{T P+T N}{T P+T N+ F P +F N}
\label{eq6}
\end{equation}

Use \eqref{eq7} to calculate the precision rate, the proportion of samples with correct, true values among the samples predicted to be correct, to reflect the accuracy of the model prediction.
\begin{equation}
Precision=\frac{T P}{T P+F P}
\label{eq7}
\end{equation}
\begin{equation}
Recall=\frac{T P}{T P+F N}
\label{eq8}
\end{equation}

Use \eqref{eq8} to calculate the recall rate, the number of positive samples are found in the data for which all true values are predicted correctly, to reflect the comprehensiveness of the model prediction.
\begin{equation}
F1=2\frac{Precision \times Recall}{Precision + Recall} = \frac{2TP}{2TP+FP+FN}
\label{eq9}
\end{equation}

The definition of the F1 score formula for each category is shown in \eqref{eq9}. F1 score can solve the balance between precision and Recall, and the higher the value, the better.

\subsection{Implementation Details}
\begin{table}[h]
\caption{Experimental setting.}
\centering
\setlength{\tabcolsep}{7mm}{
\begin{tabular}{l|l}
\hline
training config        & 224×224         \\ \hline
optimizer              & AdamW           \\ 
drop path rate         & 0               \\ 
base learning rate     & 1e-4            \\ 
min learning rate      & 1e-6            \\ 
weight decay           & 0.01            \\ 
optimizer momentum     & $\beta1$,$\beta$2=0.9,0.999 \\ 
batch size             & 32              \\ 
training epochs        & 100             \\ 
learning rate schedule & CosineAnnealing \\ 
warm up schedu         & linear               \\ 
warm up epochs         & 1               \\\hline
\end{tabular}}
\label{setting}
\end{table}

We selected the SOTA classification model of open-source with similar parameters to HiFuse for comparative experiments [9-12,17,42,43]. We implement our PyTorch-based approach by training on an NVIDIA RTX 3090 GPU with 24 GB of video memory. The base learning rate value is 1e-4, the batch size is 32, the training epoch is 100, and the cosine annealing learning rate strategy is adopted. To ensure the fairness of the experiments, we use an image size of 224×224, share the same operating environment and hyperparameters, and use the same training, validation, and test sets according to previous literature. We were conducting experiments under the mmcv\cite{mmcv} framework. 
We use softmax as the output layer and use the categorical cross-entropy loss function to calculate the loss value:
\begin{equation}
CrossEntropyLoss=-\frac{1}{N} \sum_{n=1}^N \sum_{i=1}^k y_i^t \log y_i^p
\end{equation}
where \(N\) represents the total number of samples, \(K\) represents the number of categories, \(y_i^t\) is the target label, and \(y_i^p\) is the model's predicted value output. More parameter settings are shown in Table \ref{setting}.

\subsection{Ablation Study}

\begin{table}[h]
\caption{Ablation experiment results on ISIC2018 dataset.}
\centering
\begin{tabular}{c|c|c|c|c}
\hline
\textbf{}               & \textbf{Acc\%} & \textbf{F1\%} & \textbf{Prec\%} & \textbf{Recall\%} \\ \hline
Local Path              & 77.12          & 54.12         & 60.03           & 52.77             \\ 
+ Global Block          & 79.59          & 64.32         & 64.24           & 64.66             \\ 
+ Channel\&Spatial Attn & 80.85          & 66.26         & 69.20           & 64.60             \\ 
+ IRMLP                 & 81.32          & 69.53         & 72.20           & 68.28             \\ 
+Shortcut (HiFuse-Tiny)      & 82.99          & 72.99         & 73.67           & 72.87            \\ \hline
\end{tabular}
\label{ablation}
\end{table}

As shown in the Table \ref{ablation}, we evaluated the impact of each component on the model on the ISIC2018 dataset, starting from the local path, adding the global path, channel \& spatial attention, Inverted residual MLP, and shortcut, finally, to form the final HiFuse-Tiny model, after adding the global block Acc and F1 increased by 2.47\% and 10.2\%. After adding components in HFFblock, Acc and F1 increased by 7.4\% and 8.67\%. It can also be seen that combining the global features can significantly improve the representation ability of the model, and the HFF block can provide a better fusion of global-local features. The above combination achieves an Acc of 82.99\% and an F1 value of 72.99\%.

\subsection{Visual Inspection of HiFuse}

\begin{figure}[t]
\centering
\includegraphics[width=6 in]{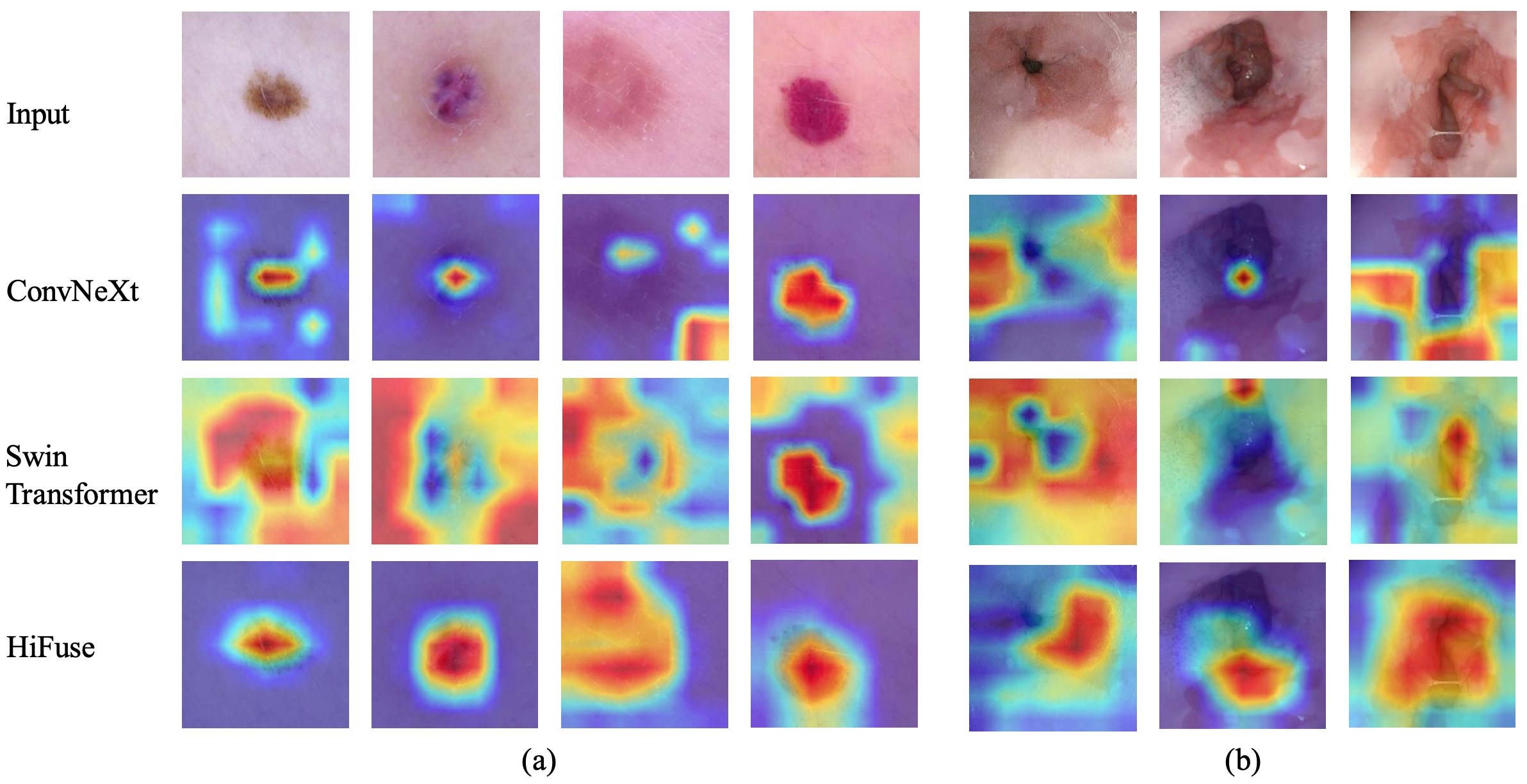}
\caption{Grad-CAM visual results of ablation experiments.}
\label{fig4}
\end{figure}

To further illustrate that our HiFuse model can effectively capture feature information of medical images, we choose the recent hierarchical structure models ConvNeX-T, Swin Transformer-T, and compare them with our HiFuse-T in this section. We adopt the method of Grad-CAM \cite{Selvaraju2017} to visualize the last layer in the model except for the linear layer and reflect the area of interest in the model in the form of a heat map. Figure \ref{fig4} (a) and (b) show the Grad-CAM visualization results of some dermoscopy and upper gastrointestinal endoscopy.

As what can be seen, ConvNeXt attaches great importance to local features, while Swin Transformer is better at paying attention to global features. The HiFuse model reflects a higher thermal value in the lesion area and more accurately covers the lesion area. Such observations demonstrate that the HiFuse can better integrate global-local features at different levels and helps the model to learn more discriminative features to pay more attention to the lesion area.

\begin{table}[t]
\caption{Performance comparison of the ISIC2018 dataset.}
\centering
\begin{tabular}{c|c|c|c|c|c|c}
\hline
\textbf{Method}        & \textbf{Params(M)} & \textbf{Flops(G)} & \textbf{Acc\%} & \textbf{F1\%}  & \textbf{Prec\%} & \textbf{Recall\%} \\ \hline
ConvNeXt-B             & 88.59              & 15.36             & 76.52          & 50.94          & 66.84           & 50.52             \\ 
VGG-19                 & 143.68             & 19.67             & 79.25          & 61.83          & 63.71           & 60.89             \\ 
Mixer-L/16             & 208.20             & 44.57             & 78.92          & 59.88          & 61.36           & 59.16             \\ 
T2T-ViT\_t-24          & 64.00              & 12.69             & 77.59          & 57.21          & 59.60           & 55.94             \\ 
DeiT-base              & 86.57              & 16.86             & 72.31          & 41.01          & 47.19           & 44.09             \\ 
VIT-B/16               & 86.86              & 33.03             & 78.32          & 60.93          & 64.16           & 60.52             \\ 
Swin-B                 & 87.77              & 15.14             & 79.79          & 63.95          & 65.09           & 63.65             \\ 
VIT-B/32               & 88.30              & 8.56              & 77.92          & 57.52          & 58.74           & 56.90             \\ 
Conformer-base-p16     & 83.29              & 22.89             & 82.66          & 72.44          & 73.31           & 71.66             \\ \hline
\textbf{HiFuse\_Tiny}  & 82.49              & 8.13              & 82.99          & 72.99          & 73.67           & 72.87             \\ 
\textbf{HiFuse\_Small} & 93.82              & 8.84              & 83.59          & 72.70          & 72.70           & 73.14          \\ 
\textbf{HiFuse\_Base}  & 127.80             & 10.97             & \textbf{84.12} & \textbf{75.32} & \textbf{76.52}  & \textbf{74.74}    \\ \hline
\end{tabular}
\label{ISIC}
\end{table}
\subsection{Results on ISIC2018 Dataset}

Table \ref{ISIC} shows the evaluation results of the proposed model and some advanced classification algorithms on the ISIC2018 dataset. We adopt the data partitioning method in the literature \cite{Cheng2022} and use the initialization network for training.

As what can be seen from the table, HiFuse has significant advantages in medical image classification. Similar to ConvNeXt and Swin Transformer, we build a hierarchical structure to improve the feature representation ability of neural networks of different scales, but 7.6\% improves the classification accuracy of our network compared with ConvNeXt and Swin Transformer, respectively (76.52\% vs. 84.12\% ) and 4.33\% (79.79\% vs. 84.12\%). Like Conformer, a multi-branch structure combines the advantages of CNN and Transformer. However, the difference lies in that we do not interact the information of the branches but fuse the branches of the HFF block at different levels to reduce the computational complexity while improving the accuracy of medical image classification.

These experimental results further confirm that the hierarchical fusion of feature information of different branches can reduce the computational complexity and improve the classification model's performance to varying degrees. HiFuse has the above two characteristics. The Flops of HiFuse-Base is only 10.97 G, which has the best classification accuracy (84.12\%) on the ISIC2018 dataset.

\subsection{Results on Covid19 Dataset}

\begin{table}[ht]
\caption{Performance comparison of the Covid19 dataset.}
\centering
\begin{tabular}{c|c|c|c|c}
\hline
\textbf{Method}        & \textbf{Acc\%} & \textbf{F1\%}  & \textbf{Prec\%} & \textbf{Recall\%} \\ \hline
ConvNeXt-B             & 55.38          & 54.68          & 54.95           & 54.81             \\ 
VGG-19                 & 59.14          & 57.55          & 59.04           & 58.13             \\ 
Mixer-L/16             & 70.43          & 70.12          & 70.38           & 70.06             \\ 
T2T-ViT\_t-24          & 63.44          & 60.34          & 65.68           & 61.89             \\ 
DeiT-base              & 50.54          & 39.31          & 44.47           & 47.96             \\ 
VIT-B/16               & 65.05          & 64.88          & 64.90           & 64.87             \\
Swin-B                 & 60.75          & 56.36          & 63.20           & 58.95             \\ 
VIT-B/32               & 61.83          & 60.59          & 61.89           & 60.94             \\ 
Conformer-base-p16     & 75.81          & 75.60          & 76.81           & 77.81             \\ \hline
\textbf{HiFuse\_Tiny}  & 74.73          & 74.67          & 74.65           & 74.73             \\ 
\textbf{HiFuse\_Small} & \textbf{76.88} & \textbf{76.31} & \textbf{77.78}  & \textbf{76.19}    \\ 
\textbf{HiFuse\_Base}  & 76.34          & 76.17          & 76.30           & 76.11             \\ \hline
\end{tabular}
\label{Covid}
\end{table}

Table \ref{Covid} shows the evaluation results of the COVID19-CT dataset. We adopt the literature [39] data partitioning method and use the initialization network for training. Among them, the bold represents the best performance.

DeiT\cite{Touvron2021} has limited classification performance on this dataset. Our analysis for this situation may be that the network structure is mainly designed for large datasets (such as ImageNet) and is not suitable for small datasets such as COVID19-CT. It can be seen from the experimental results that the pure convolutional models ConvNeXt\cite{Liu2022} and VGG\cite{Simonyan2014} are not good at extracting features from small-sample CT datasets, while HiFuse and Conformer using multi-branch network structures have higher classification performance.

HiFuse-Small's accuracy (76.88\%) and F1 value (76.31\%) achieved the best performance on this dataset. These results demonstrate that HiFuse can be robust to other applications in the same domain and maintain high classification performance.

\subsection{Results on Kvasir Dataset}
\begin{table}[ht]
\caption{Performance comparison of the Kvasir dataset.}
\centering
\begin{tabular}{c|c|c|c|c}
\hline
\textbf{Method}        & \textbf{Acc\%} & \textbf{F1\%}  & \textbf{Prec\%} & \textbf{Recall\%} \\ \hline
ConvNeXt-B             & 74.60          & 74.61          & 74.78           & 74.64             \\ 
VGG-19                 & 77.75          & 77.75          & 77.86           & 77.83             \\ 
Mixer-L/16             & 74.30          & 74.14          & 74.43           & 74.34             \\ 
T2T-ViT\_t-24          & 76.90          & 76.78          & 77.60           & 76.91             \\ 
DeiT-base              & 52.15          & 48.48          & 56.72           & 52.29             \\ 
VIT-B/16               & 76.10          & 75.94          & 76.49           & 76.23             \\ 
Swin-B                 & 77.30          & 77.29          & 77.74           & 77.44             \\ 
VIT-B/32               & 73.80          & 73.50          & 74.24           & 73.72             \\ 
Conformer-base-p16     & 84.25          & 84.27          & 84.45           & 84.37             \\ \hline
\textbf{HiFuse\_Tiny}  & 84.85          & 84.89          & 84.96           & 84.90             \\ 
\textbf{HiFuse\_Small} & \textbf{85.00} & \textbf{84.96} & \textbf{85.08}  & \textbf{85.00}    \\ 
\textbf{HiFuse\_Base}  & 84.35          & 84.41          & 84.50           & 84.48             \\ \hline
\end{tabular}
\end{table}
In order to further explore the generalization ability of HiFuse, we conduct experiments in the Kvasir dataset according to the division method and 2-fold cross-validation in the literature\cite{pogorelov2017kvasir}, and the final results are averaged.

It can be seen that HiFuse-Small's accuracy (85\%) and F1 value (84.96\%) achieve the best performance in this dataset. Moreover, the accuracy of HiFuse-Base decreases slightly due to the increase in depth, which can be improved slightly by setting the appropriate drop path hyperparameter (the table ensures a fair comparison, not shown). This reminds us that HiFuse-Small with lower depth may have better classification performance when HiFuse is applied to other smaller medical image classification datasets.

In short, capturing global-local features has obvious advantages in medical image classification, and our HiFuse can achieve higher accuracy with lower computational complexity.

\section{Discussion}
Compared with natural images, medical images have diverse characteristics \cite{Zhou2021}, fewer data, and different medical equipment, resulting in limited training data and models that cannot focus well on classification features. Many models that perform well in natural image classification tasks are used in medical image field; satisfactory results cannot be obtained. Therefore, finding an efficient and robust backbone network for medical images remains challenging.

Compared with the convolution operation, Transformer's self-attention has a global receptive field to mine the long dependencies between pixels, and has a stronger generalization ability. Extensive experiments show that local spatial features are equally important in medical image processing. Based on the above requirements, we design a three-branch hierarchical feature fusion model HiFuse, which fuses global-local feature representations of different scales through the HFF block. In ISIC2018, Covid-19, and Kvasir medical classification datasets, comparison experiments have been conducted to verify that HiFuse has the best results. In addition, our HiFuse benefits from a hierarchical structure with rich scalability, linear computational complexity, and a wider range of applications.

Although our model focused on the task of medical image classification, the ideas presented in this paper can provide researchers with some new ideas for global and local feature fusion. The hierarchical fusion method in the architecture is designed to be easily extended and upgraded. 

Our model can be further improved in future research:

1. According to the specific situation of the task, assign the network depth and width of different branches to make the local features and global representation more directional.

2. Design targeted dynamic hierarchical feature selection for different datasets to improve the performance of HiFuse further.

3. Continue HiFuse research in medical image segmentation and multimodal.

\section{Conclusion}

In this paper, we propose HiFuse, a three-branch hierarchical fusion classification model, and the modular design has rich scalability and linear computational complexity. In HiFuse, the local feature block extracts local features, the global feature block captures global representations, and the hierarchical feature fusion block (HFF block) fuses local features and global representations at different scales, which can comprehensively mine the deep-shallow features and global-local features of the lesion area in the medical image classification task. Experiments show that our proposed method achieves good results on three medical image datasets. We believe this work can contribute to various downstream tasks in medical imagery.

\bibliographystyle{unsrtnat}
\bibliography{ZZZref}  






\end{document}